\documentclass{aa520}  
\usepackage{graphicx}

\begin{document}

\title{On the CMB large-scale angular correlations} 

   \author{
Armando Bernui\inst{},\thanks{
on leave from Universidad Nacional de Ingenier\'{\i}a, 
Facultad de Ciencias, Apartado 31 - 139,\, Lima 31 -- Peru}
Thyrso Villela\inst{},
Carlos A. Wuensche\inst{},
Rodrigo Leonardi\inst{},
\and
Ivan Ferreira\inst{}
          }

   \offprints{A. Bernui}

   \institute{Instituto Nacional de Pesquisas Espaciais, 
   Divis\~{a}o de Astrof\'{\i}sica, Av. dos Astronautas 1758, 
   12227-010, S\~ao Jos\'e dos Campos, SP -- Brazil \\
\email{bernui@das.inpe.br}   \\
\email{thyrso@das.inpe.br}   \\  
\email{alex@das.inpe.br}     \\
\email{leonardi@das.inpe.br} \\
\email{ivan@das.inpe.br}
             }

   \date{Received / Accepted}
\abstract{
{We study the large-scale angular correlation signatures of the Cosmic 
Microwave Background (CMB) temperature fluctuations from WMAP data
in several spherical cap regions of the celestial sphere, 
outside the Kp0 or Kp2 cut-sky masks.}
{We applied a recently proposed method to CMB temperature maps, which 
permits an accurate analysis of their angular correlations in the 
celestial sphere through the use of normalized histograms of the number 
of pairs of such objects with a given angular separation versus their 
angular separation.
The method allows for a better comparison of the results from observational 
data with the expected CMB angular correlations of a statistically isotropic 
Universe, computed from Monte Carlo maps according to the WMAP best-fit 
$\Lambda$CDM model.}
{We found that the, already known, anomalous lack of large-scale power in 
full-sky CMB maps are mainly due to missing angular correlations of 
quadrupole-like signature. 
This result is robust with respect to frequency CMB maps and cut-sky masks. 
Moreover, we also confirm previous results regarding the unevenly distribution
in the sky of the large-scale power of WMAP data. 
In a bin-to-bin correlations analyses, measured by the full covariance matrix 
$\chi^2$ statistic, we found that the angular correlations signatures in 
opposite Galactic hemispheres are anamalous at the 98\%--99\% confidence 
level.}}

\titlerunning{On the CMB large-scale angular correlations}
\authorrunning{Bernui et al.}

\maketitle


\section{Introduction}

In recent years, after the COBE mission, experiments of observational 
cosmology evolved both in the accuracy  and in the angular resolution 
of the Cosmic Microwave Background (CMB) measurements (see, e.g. 
Bersanelli et al.~\cite{Bersanelli} and references therein). 
However, such experiments were designed to map small patches of the 
celestial sphere and therefore the study of its properties 
concerning the large-scale angular correlations was not possible.

This situation changed with the Wilkinson Microwave Anisotropy Probe 
(WMAP), which produced full-sky CMB maps in five frequency bands 
(termed K, Ka, Q, V, and W maps), although containing different 
amounts of contaminations from our galaxy (Bennett et al.~\cite{WMAPa}).
This superb quality data set motivated a number of detailed studies 
concerning the CMB properties. 
As a result, several recent works have reported claims regarding some 
anomalies found in these data. 
These anomalies include the low-order multipole values (Bennett et 
al.~\cite{WMAPa}; Efstathiou~\cite{GE}; Tegmark et al.~\cite{TOH}; 
Gazta\~naga et al.~\cite{Gaztanaga}; Slozar et al.~\cite{Slosar}; 
O'Dwyer et al.~\cite{ODwyer}), 
the alignment of some low-order multipoles 
(Tegmark et al.~\cite{TOH}; de Oliveira-Costa et al.~\cite{OTZH}; 
Eriksen et al.~\cite{Eriksen04b}; Bielewicz et al.~\cite{Bielewicz05}; 
Land and Magueijo~\cite{Land05}), and 
an unexpected asymmetric distribution on the sky of the large-scale power 
of CMB data (Eriksen et al.~\cite{Eriksen04a},~\cite{Eriksen05a}; Hansen 
et al.~\cite{Hansen04a},~\cite{Hansen04b},~\cite{Hansen04c}; 
Jaffe et al.~\cite{Jaffe}; Schwarz et al.~\cite{Schwarz}; Copi et 
al.~\cite{Copi05}). 

In this work we focus on the study of the large-scale angular 
correlations (hereafter termed Angular Correlations Signatures ACS)
of the CMB temperature maps from WMAP data.
Since the vast majority of the cosmological information is contained in 
the two-point temperature correlation function, it seems natural to 
study the ACS by looking at such function (actually, an equivalent 
of it). 
Our analyses of the ACS in CMB maps, for data outside the region defined 
by the Kp0 or Kp2 WMAP masks, allow to perform a close inspection of two 
interesting issues. 
First, we investigate --through a full-sky map analysis-- the large-scale 
power in WMAP data, and its connection with the low quadrupole moment 
issue.
Second, we study --through a partial-sky coverage analysis-- 
the possible uneven distribution on the celestial sphere of the CMB 
large-scale power.
The significance of our results are evaluated by comparing these 
results against those obtained from Monte Carlo CMB maps produced with 
the WMAP best-fitting $\Lambda$CDM model properties (Hinshaw et 
al.~\cite{Hinshaw03b}).
Finally, using a $\chi^2$ statistics we assess the confidence level of 
the North/South asymmetry in WMAP data compared to 1\,000 Monte Carlo 
realizations of CMB skies.

\section{Analyses method}

Recently, Bernui \& Villela (\cite{BV}) introduced a new method to study 
the large-scale ACS in the distribution of cosmic objects in the sky.
This method, called the Pair Angular Separation Histogram (PASH) method,  
consists in first calculating the angular distances between all pairs of 
cosmic objects, listed in a catalog, and then constructing the normalized 
histogram of the number of pairs with a given angular separation versus 
their angular separation. 
A catalog with a large number of objects can be divided in a number (say 
$K$) of comparable sub-catalogs (i.e. ensembles of iso-number of objects 
sharing analogous physical properties). 
After that, one computes a PASH with each one of these sub-catalogs and 
average them to obtain the Mean-PASH (MPASH).
The difference between the MPASH, caculated using a observational catalog, 
and the Expected-PASH (EPASH), obtained assuming the statistical isotropy 
hypothesis (hereafter the MPASH-minus-EPASH function), shows the ACS of 
the cosmic objects in such a catalog. 

In the CMB temperature maps the celestial sphere is partitioned in a set 
of equal-area pixels, where to each pixel is assigned a weighted CMB 
temperature. 
To obtain the ACS of a given map, one divides the set of pixels in $K=2$ 
sub-maps, one for the negative CMB temperature fluctuations and the other 
for the positive ones, and proceeds as before averaging $(K=)$ 2 PASHs.
If, for computational problems, the number of pixels in these sub-maps is 
too large, one can divide them in, say $K_{-}$ and $K_{+}$, 
sub-maps (with iso-number of pixels of analogous temperatures), 
and proceeds to compute the MPASH averaging the $K = K_{-} + K_{+}$ PASHs,  
and finally, one plots the MPASH-minus-EPASH function. 
It can be shown  (see Bernui~\cite{AB}) that the MPASH-minus-EPASH function 
is independent of the number of sub-maps $K_{-} \ge 1$ and $K_{+} \ge 1$ in 
which the original CMB map is divided. 

The PASH method is also applicable to incomplete sky maps, including 
disconnected regions such as those resulting from the application of the 
WMAP masks. 
This method is similar in philosophy to the two-point temperature 
correlation function, except that the PASH method has zero mean, 
because the MPASH and EPASH are each one normalized histograms. 
This simple fact allows us to achieve a deeper study of the ACS in CMB 
temperature maps, like: 
the intensity of all their relative mimima and maxima and their 
corresponding angular scales, their intersections with the horizontal 
axis (i.e. the zeros of the function), etc; it also leads to 
calculate the variance $\sigma^2$ of the MPASH-minus-EPASH function $f_i$ 
(which has zero mean, i.e., $\,\sum_{i=1}^{N_{\rm{bins}}} \, f_i = 0$) 
\vspace{-0.9mm}
\begin{eqnarray}
\sigma^2 \,=\, \frac{1}{N_{\rm{bins}}} \sum_{i=1}^{N_{\rm{bins}}}\, 
f_i^{\,2} \,\, . \nonumber
\end{eqnarray}

\section{CMB data analyses}

In this section we present the ACS of the WMAP data. 
For these analyses, we used six CMB maps: the individual Q-, V-, and W-band
data produced from the (Q$_1$,Q$_2$), (V$_1$,V$_2$), and (W$_1$,W$_2$,W$_3$,
W$_4$) differential assemblies (DA), respectively, as provided on 
LAMBDA~\footnote{
\mbox{$\!$http:$\!\backslash\!\backslash$lambda.gsfc.nasa.gov
$\!\!\backslash$product$\!\backslash$map$\!\backslash$current}
$\!\backslash$IMaps$_{-}$cleaned.cfm} 
(Bennett et al.~\cite{WMAPb}), 
the {\sc Coadded} WMAP map (which combines the eight DA in the Q-, $\!$V-, 
$\!$and 
W-bands using the inverse-variance noise weights as described by Hinshaw et 
al.~(\cite{Hinshaw03b})), 
the {\em cleaned} CMB map (Tegmark et al.~\cite{TOH}, hereafter the TOH 
map), and the Lagrange Internal Linear Combination (Eriksen et 
al.~\cite{Eriksen04b}, hereafter the LILC map). 
It is known that the last two maps are not fully reliable for quantitative 
analysis because still have problems with residual foregrounds 
(Eriksen et al.~\cite{Eriksen05b}); here we use them just as a supporting 
\mbox{evidence}. 
After applying the Kp0 or Kp2 WMAP masks in all these six maps, we remove 
their residual monopole and dipole components, and correct them for the 
dynamic \mbox{quadrupole}. 

The examination of the WMAP data performed here comprehends two types of 
sky analyses.
$\!$We want to remark that in both cases we consider data outside 
the region defined by the Kp0 or the Kp2 mask, to avoid from the 
beginning residual foreground contaminations (Bennett et al.~\cite{WMAPb}; 
Eriksen et al.~\cite{Eriksen04b},~\cite{Eriksen05b}; Bielewicz et 
al.~\cite{Bielewicz04}).
The first type is the full-sky analysis, and the second one is the partial-sky 
analysis carry out through antipodal spherical caps of $45^{\circ}$, 
$60^{\circ}$, and $90^{\circ}$ of aberture (this means that the maximum 
angular separation between pixels in each case is twice these angle values).

We use CMB maps with HEALPix (G\'orski et al.~\cite{Gorski}) resolution 
N$_{\rm{side}}=128$, which amounts to 196\,608 pixels. 
In the following figures we 
consider the bin width as $0.45^{\circ}$, and the number of bins as 
$N_{\rm{bins}}=400$; moreover, the shaded areas correspond to 1-sigma
confidence regions.

\subsection {Full-sky analyses (outside Kp2 or Kp0 masks)}

In this subsection we shall discuss the ACS of the WMAP maps, analysing the 
CMB data outside the regions defined by the Kp2 and Kp0 masks.
The objective of such analysis is to investigate previous claims regarding 
the lack of power in the large-scale angular correlations in full-sky data
(see Bennett et al.~\cite{WMAPa}; Efstathiou~\cite{GE}; Slosar et 
al.~\cite{Slosar}) and its relationship to the low-order multipole values.

First we show the ACS of the WMAP data using the Kp2 mask. 
In fact, the suitability of using each WMAP mask is well known in the 
literature, and many works claim that the less severe cut-sky Kp2 could 
be enough for cosmological purposes (large angular scales).
We have computed the ACS for both masks, although we only show the ACS 
for the Kp2 mask case, and for all maps under investigation to assert the 
robustness of our results under different sky cuts. 

\begin{figure}
\includegraphics[width=8.8cm, height=5.1cm]{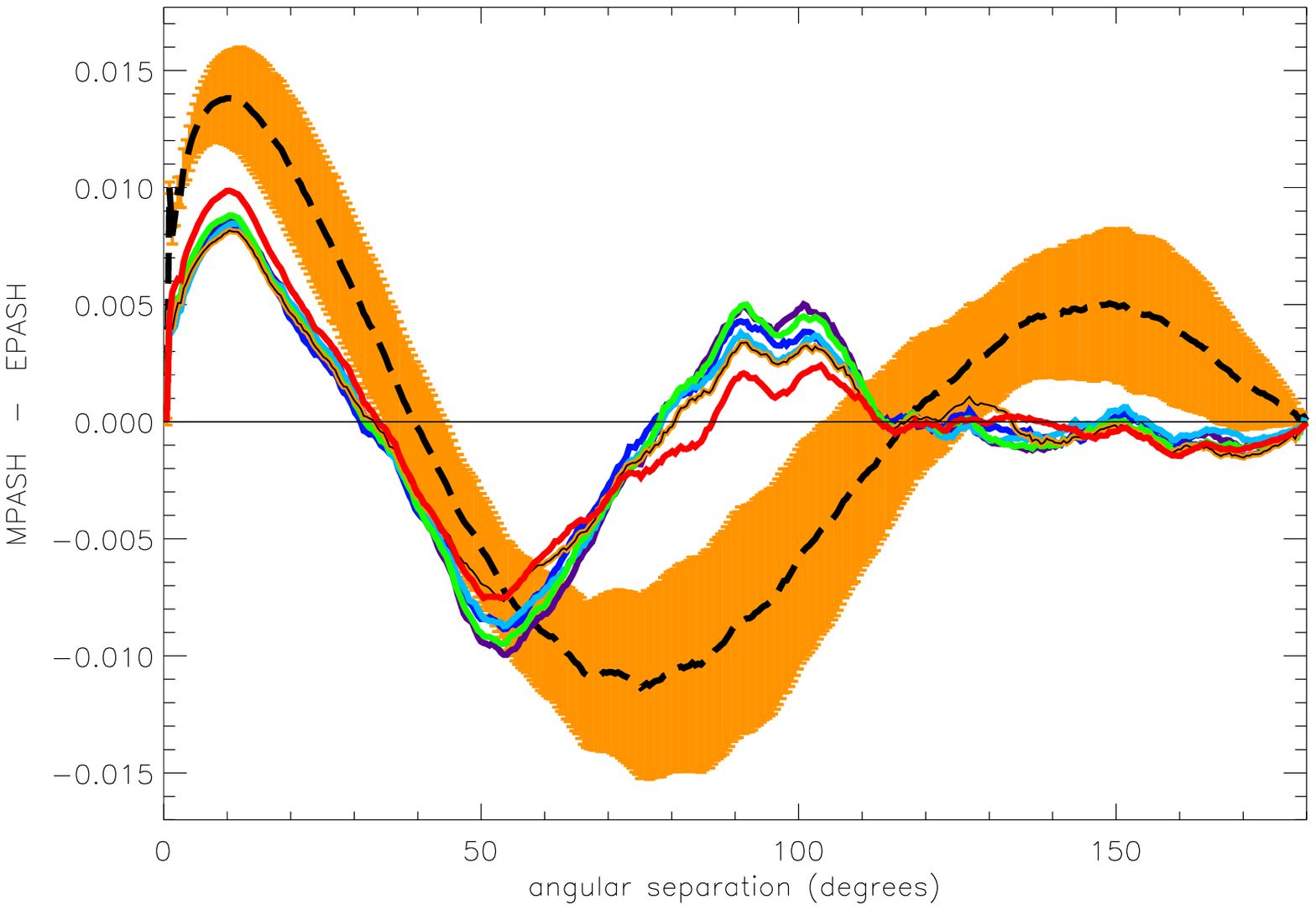}
\includegraphics[width=8.8cm, height=5.1cm]{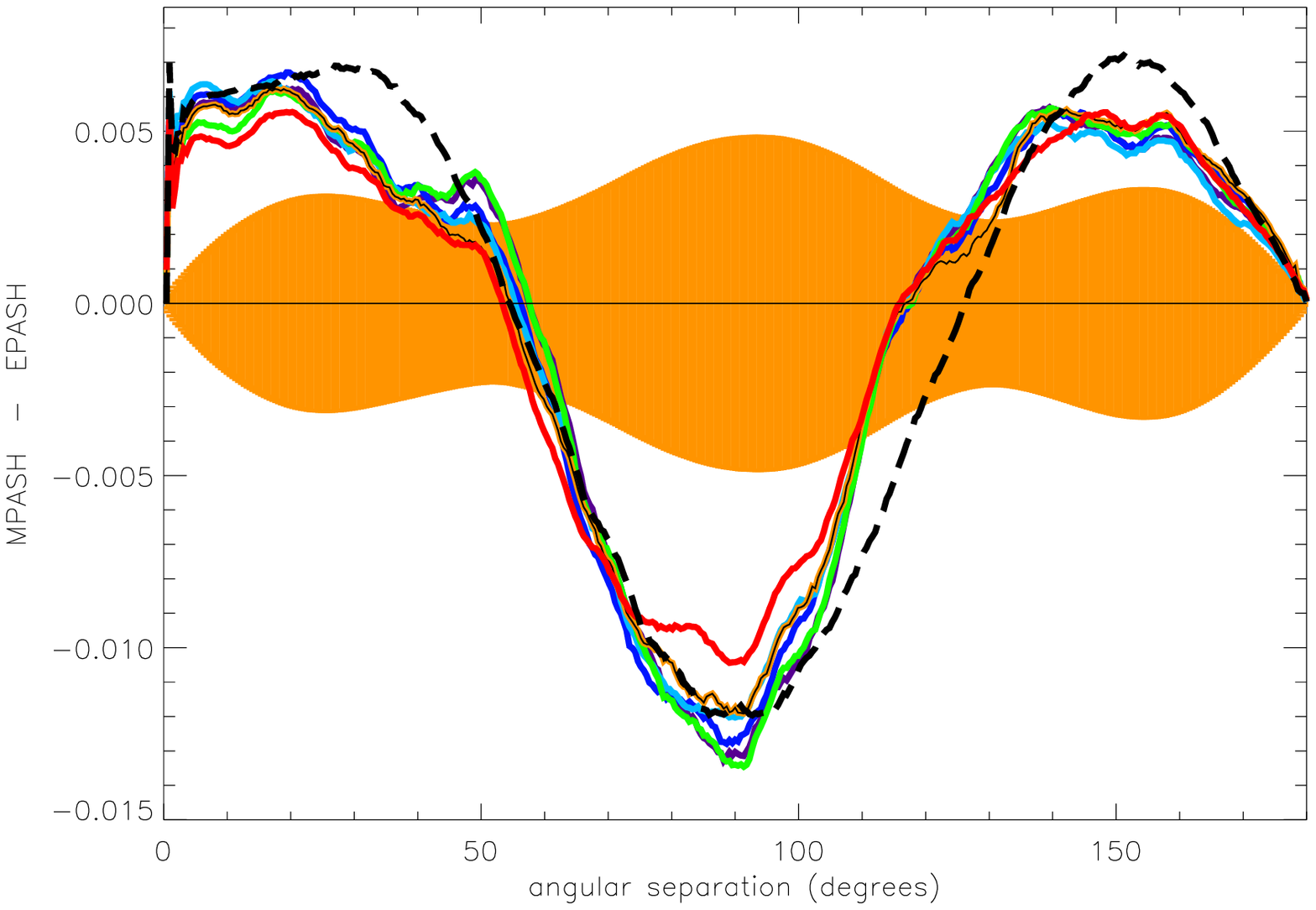}
\caption{Top panel: ACS in the six CMB maps: 
Q (violet line), V (blue line), W (light-blue line), Coadded (green line), 
TOH (brown line), and LILC (red line), after applying the Kp2 mask.
For comparison we also plot the average of 1\,000 MPASH-minus-EPASH 
functions obtained from the same number of Monte Carlo CMB maps 
(dashed line, also called the expected function). 
One unit in the vertical axis corresponds to 5\,460$\mu$K$^2$.
Bottom panel: The differences of the expected function minus the six 
MPASH-minus-EPASH functions from the top panel, where we observe that 
these ACS correspond mainly to a quadrupole-like signature. 
The dashed line here corresponds to the MPASH-minus-EPASH of a 
quadrupole (corrected for the Kp2 mask) with $C_2 \simeq 670 \mu$K$^2$. 
Notice that these curves represent the missing ACS in WMAP data to fit 
the expected function, which have 
$C_2^{\Lambda\mbox{\rm\sc cdm}} = 870 \mu$K$^2$.} 
\label{figure1}
\vspace{-0.27cm}
\end{figure}

The results are reported in Figure~\ref{figure1}. 
In the top panel of Fig.~\ref{figure1} we show the ACS from the six CMB 
maps here investigated, considering data outside the region determined by 
the Kp2 mask.
For comparison we also plot, as a dashed line, the average of 1\,000 
MPASH-minus-EPASH functions (termed the {\em expected function}) 
obtained from the same number of Monte Carlo CMB maps. 
These realizations were produced using an input power spectrum generated 
by the {\sc cmbfast} code (Seljak \& Zaldarriaga~\cite{Seljak}) 
considering the WMAP best-fitting $\Lambda$CDM model properties. 
The maps were produced using the map-making code {\sc synfast} from 
HEALPix.
The disagreement observed between the observational data from WMAP and 
the expected angular correlation function is better appreciated in the 
bottom panel of Fig.~\ref{figure1}, 
where we plotted the difference of such expected function minus each 
one of the MPASH-minus-EPASH functions from the six CMB maps.  
What is revealed in these difference plots is that the ACS missing 
in CMB data at large-angular scales correspond to a quadrupole-like 
angular correlation signature (see, e.g., Bernui~\cite{AB}). 
To assess this, we plot the angular correlation function of a 
quadrupole (corrected for the Kp2 mask) with moment 
$C_2 \simeq 670 \mu$K$^2$. 
Since these curves (bottom panel Fig.~\ref{figure1}) represent 
the missing ACS in WMAP data to fit the expected function, where 
$C_2^{\Lambda\mbox{\rm \small CDM}} = 870 \mu$K$^2$, 
one finds $C_2^{\mbox{\rm \small WMAP}} \simeq 190 \pm 30 \mu$K$^2$, 
that are in good agreement with Bielewicz et al. 
(\cite{Bielewicz04}) who obtained 
$C_2 = 165 \pm 34, 216 \pm 42, 229 \pm 44, 191 \pm 38\, \mu$K$^2$, 
for the Q, V, W, and Coadded maps, respectively).
This result, which appears to be robust with respect to the frequency 
CMB map and to cut-sky mask (we also performed analyses with the Kp0 
mask obtaining similar results), proves that the lack of angular power 
previously found in the two-point temperature correlation function of 
WMAP data is mainly due to the low quadrupole moment 
(Efstathiou~\cite{GE}; Gazta\~naga et al.~\cite{Gaztanaga}; Slosar et 
al.~\cite{Slosar}).

\begin{figure}[t]
\includegraphics[width=8.8cm, height=5.1cm]{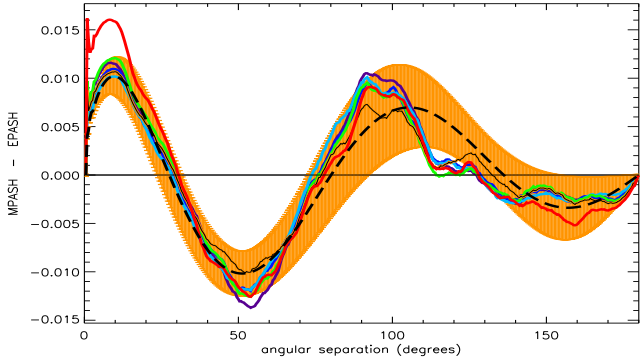}
\caption{The ACS for the six CMB maps (the nomenclature of the curves 
follows the same color's pattern as before) with the quadrupole component 
removed. 
The dashed line is the expected function for the case of zero quadrupole 
moment.} \label{figure2}
\vspace{-0.5cm}
\end{figure}

Furthermore, to confirm this result we remove the quadrupole component, 
after applying the Kp2 mask, and compute the ACS in these maps.
The result is shown in Figure~\ref{figure2} where we also plot for 
comparison the expected function for the case of zero quadrupole 
compoment. 
Again this result is robust with respect to the frequency CMB map 
analyzed.

\subsection {Partial-sky analyses}

The basic idea is to investigate different regions of the celestial sphere,
namely spherical caps of various sizes, whose vertices point in the 
direction of the North Galactic Pole (NGP) and the South Galactic Pole 
(SGP), with respect to the Galactic coordinate system.
Comparison of the ACS in antipodal caps of $45^{\circ}$, $60^{\circ}$, 
and $90^{\circ}$ of aberture allows a quantitative assessement of the 
claimed asymmetry in the distribution of the CMB power between the northern 
and southern hemispheres (Eriksen et al.~\cite{Eriksen04b}).

\begin{figure}[t]
\includegraphics[width=8.8cm, height=5.2cm]{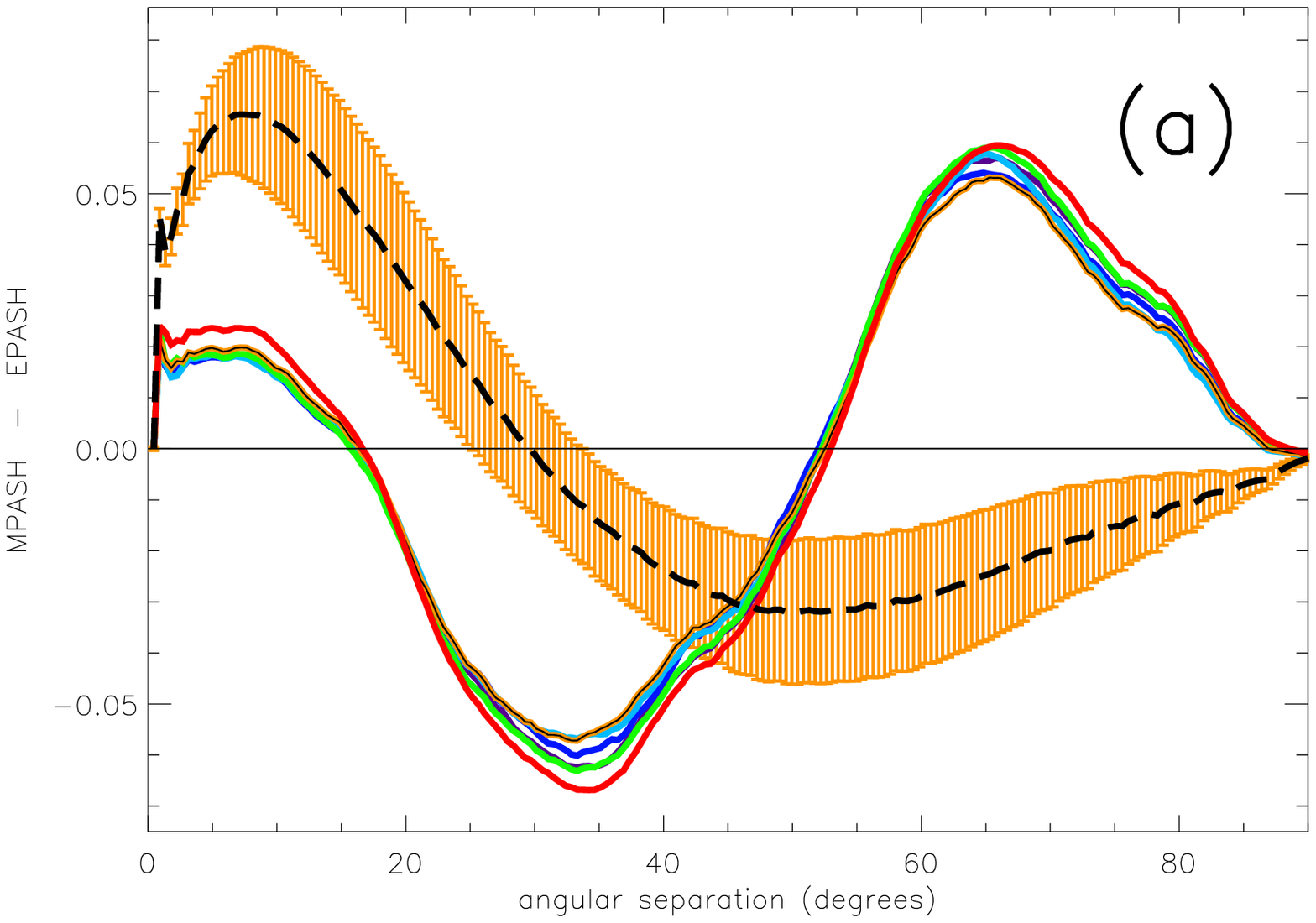}
\includegraphics[width=8.8cm, height=5.2cm]{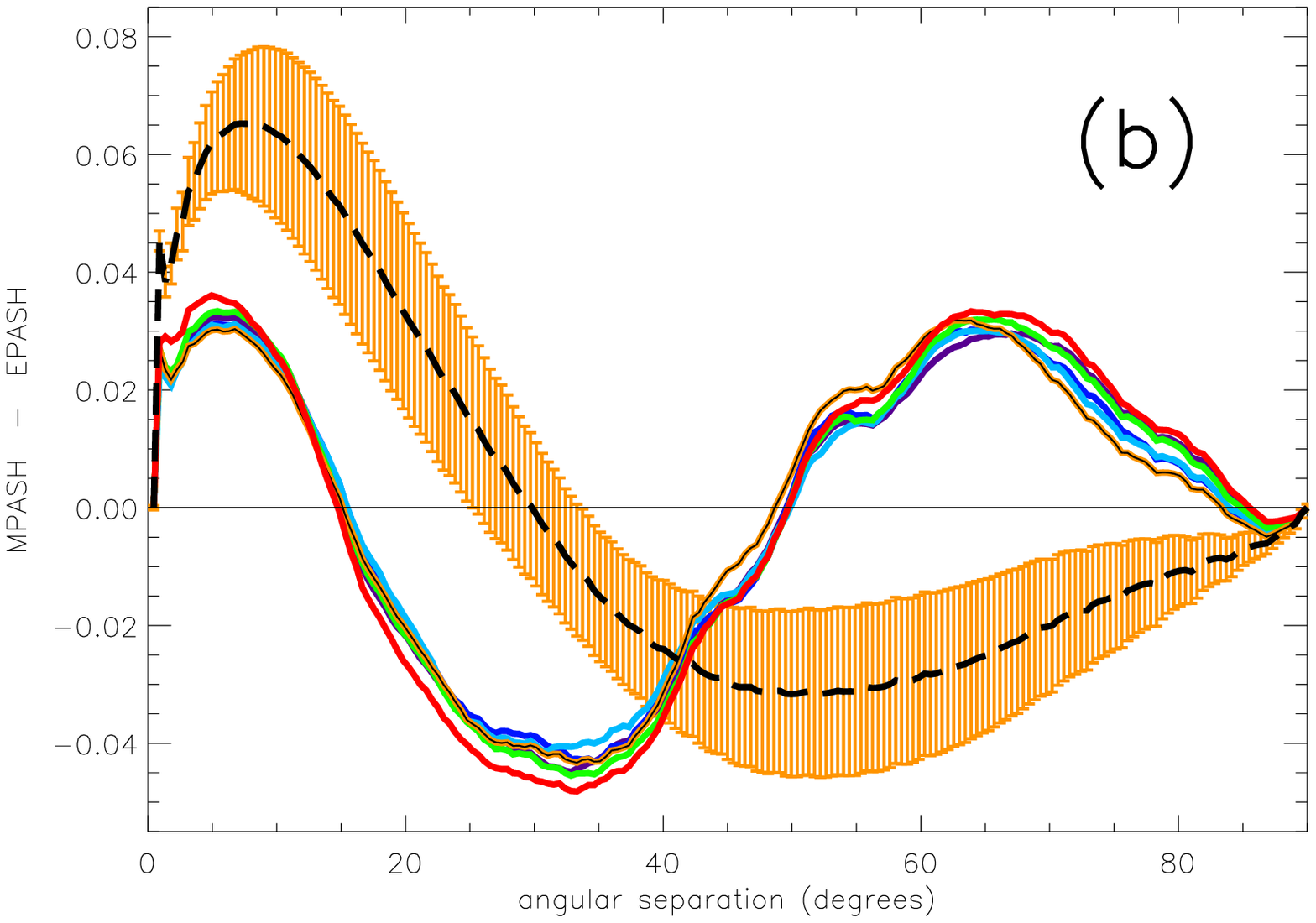}
\includegraphics[width=8.8cm, height=5.2cm]{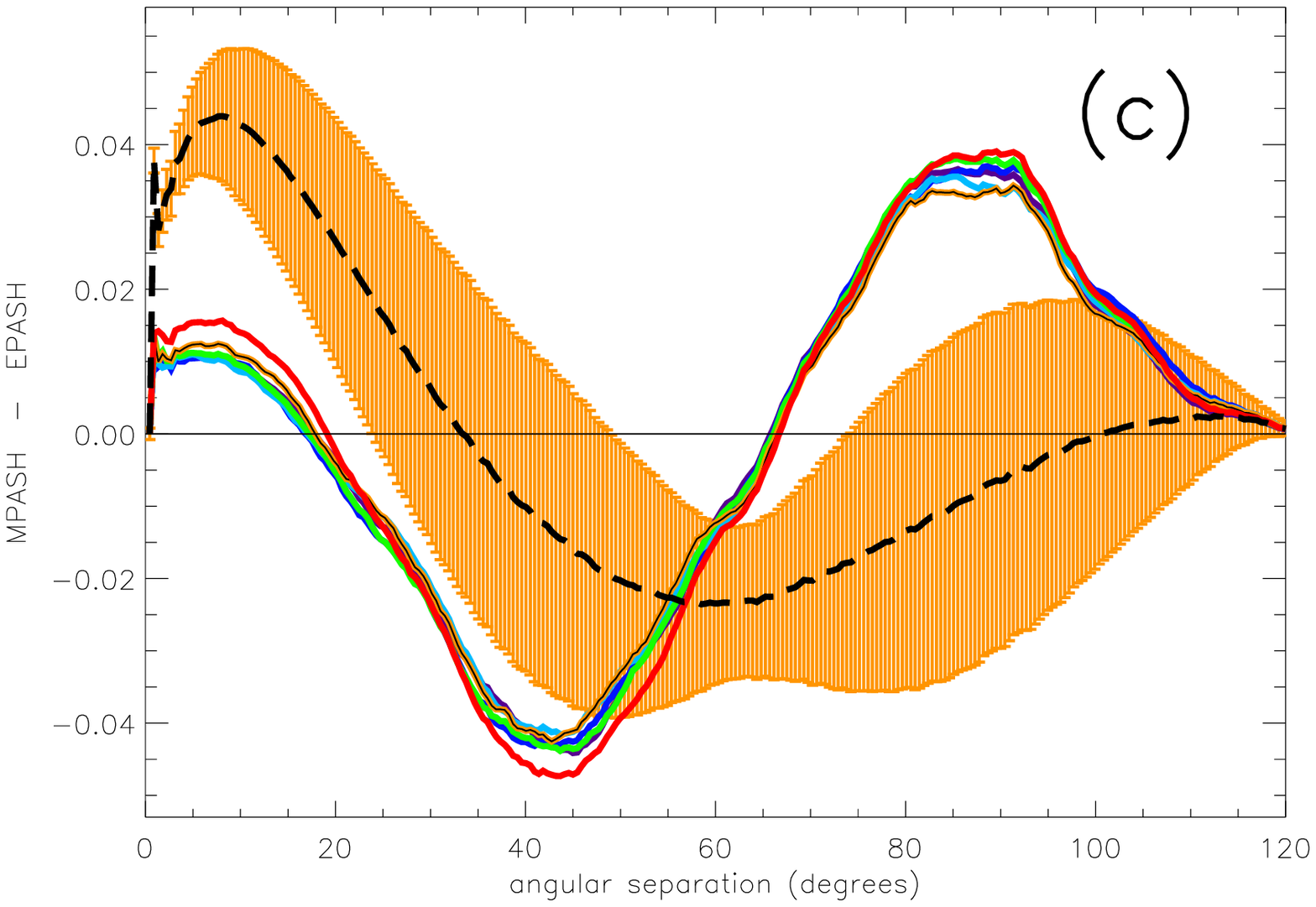}
\includegraphics[width=8.8cm, height=5.2cm]{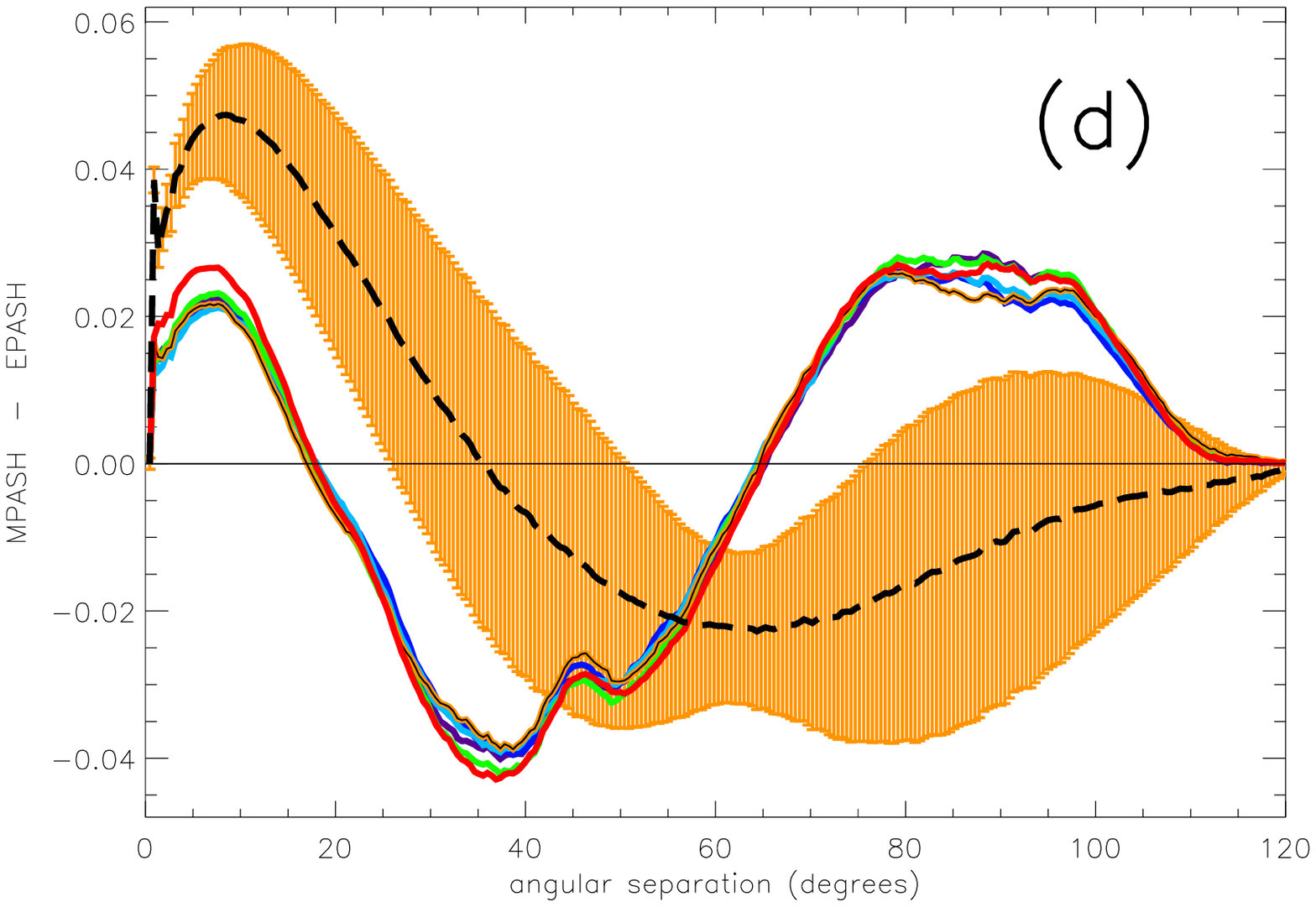}
\caption{
The ACS in the $45^{\circ}$-caps in the NGH (a) and in the SGH (b), 
and 
the ACS in the $60^{\circ}$-caps in the NGH (c) and in the SGH (d), 
from the six CMB maps (Q, V, W, Coadded, TOH, and LILC), 
after applying the Kp2 mask. 
In all plots the dashed line is the average of MPASH-minus-EPASH 
functions obtained from the corresponding sky regions in 1\,000 Monte 
Carlo maps, respectively.
The nomenclature of the curves follows the same color's pattern 
as in Figure~\ref{figure1}.
}
\label{figure3}
\end{figure}

\begin{figure}
\includegraphics[width=8.8cm, height=5.2cm]{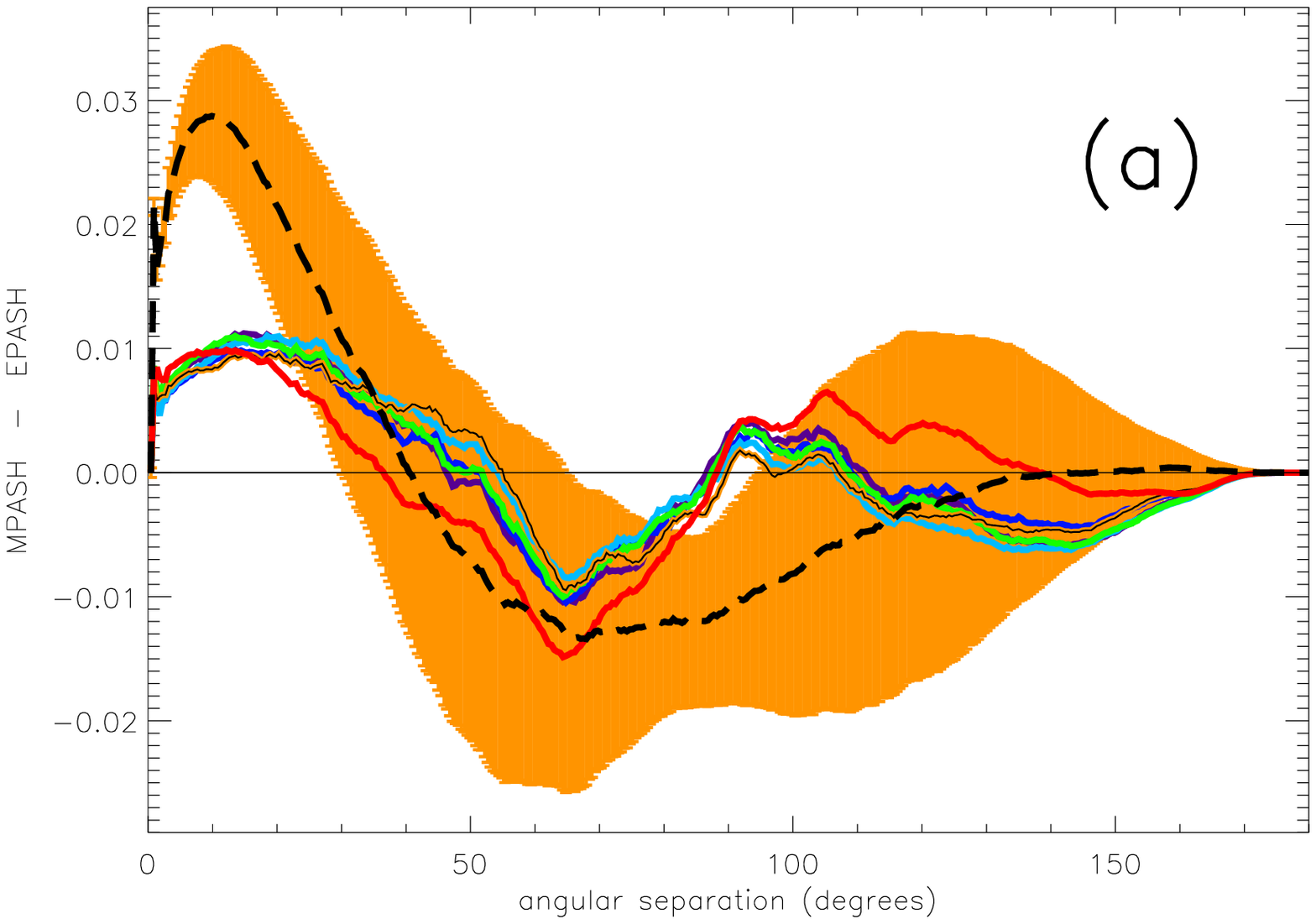}
\includegraphics[width=8.8cm, height=5.2cm]{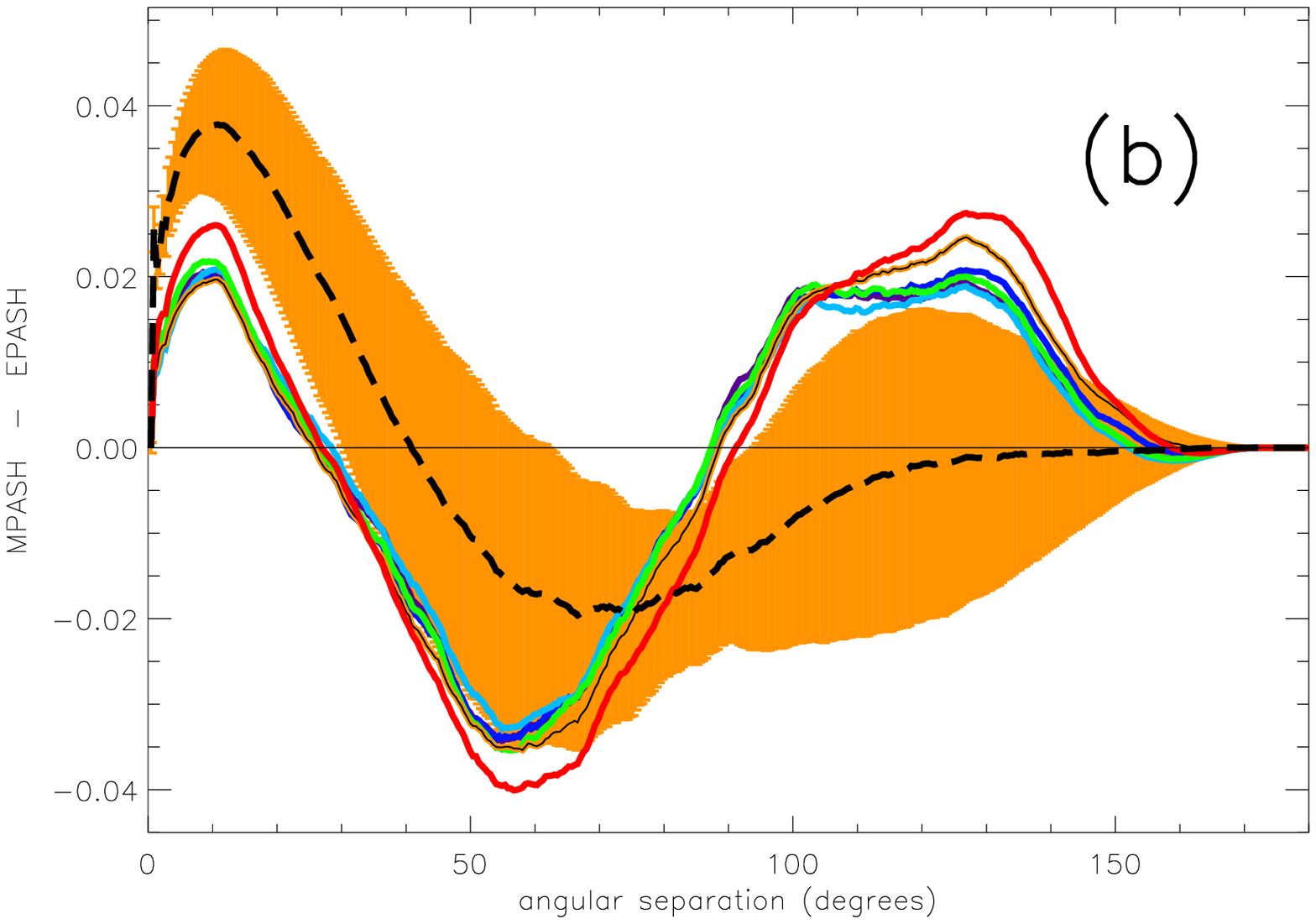}
\includegraphics[width=8.8cm, height=5.2cm]{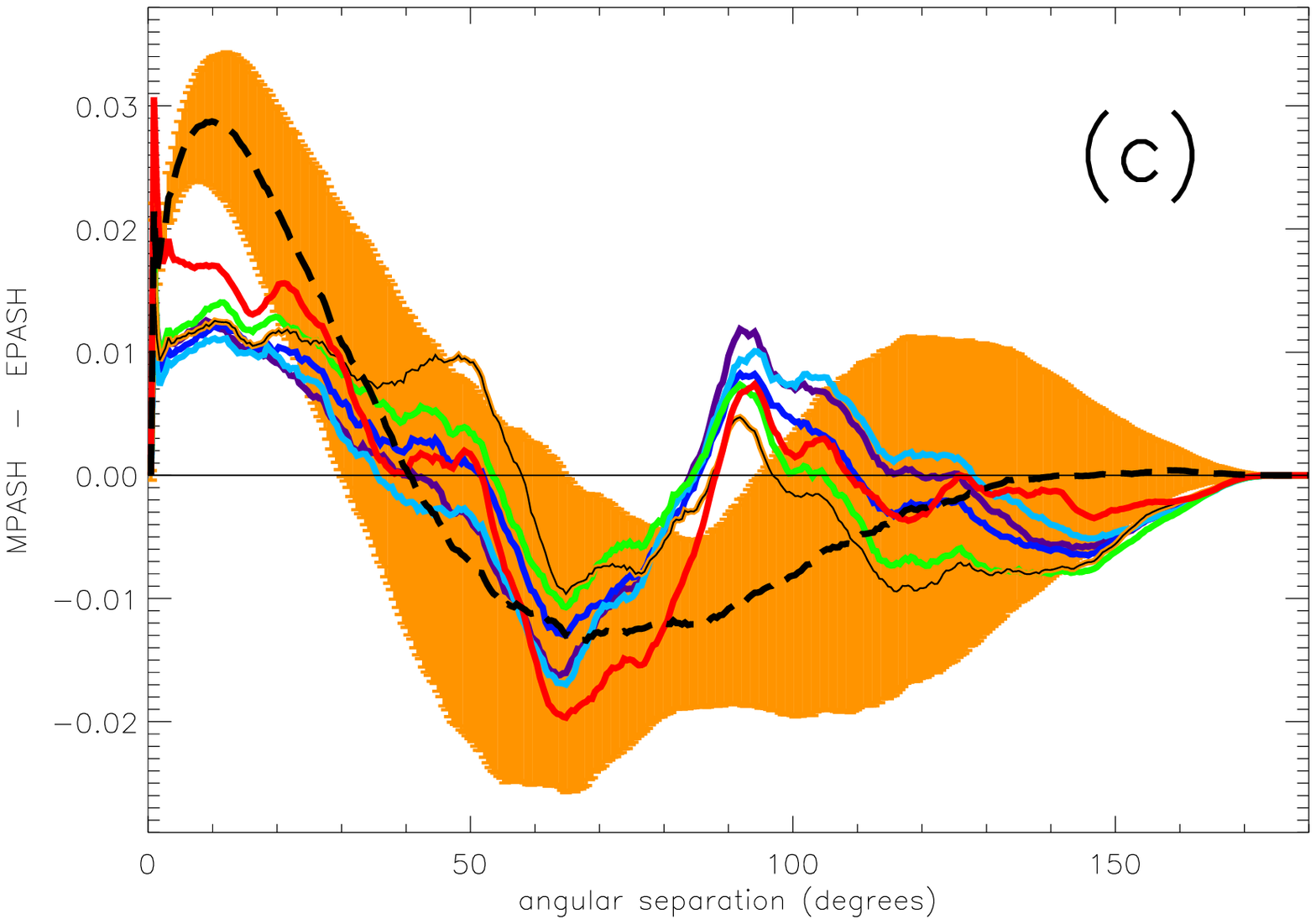}
\includegraphics[width=8.8cm, height=5.2cm]{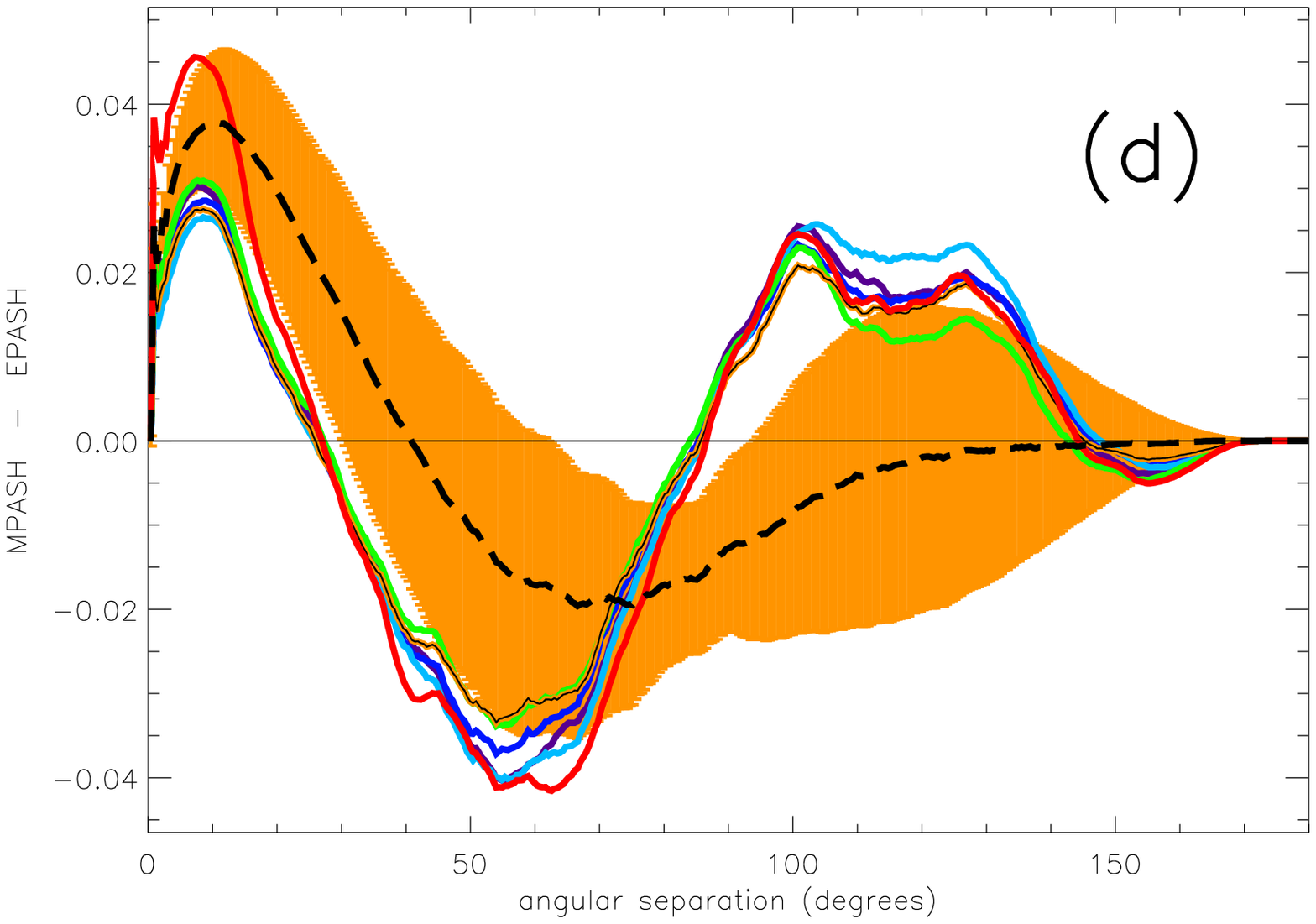}
\caption{
The ACS in the $90^{\circ}$-caps in the NGH (a) and in the SGH (b), 
from the six CMB maps (Q, V, W, Coadded, TOH, and LILC), after applying 
the Kp2 mask. 
Removing the quadrupole component, after applying the Kp2 mask, in the 
six CMB maps we obtain the ACS in the $90^{\circ}$-caps in the NGH (c), 
and in the SGH (d). 
For comparison, the dashed line and the shaded region in panel (a) 
(panel (b)) is the same as in panel (c) (panel (d)).
As before, the nomenclature of all the curves follows the same color's 
pattern as in Figure~\ref{figure1}.
}
\label{figure4}
\end{figure}

The resulting ACS are shown in Figures~\ref{figure3} and \ref{figure4}, 
where the expected functions result from data on the same patch of the sky 
from Monte Carlo CMB maps.
In Figure~\ref{figure3} we present four plots: 
panels (a) and (b) reveal the ACS in the $45^{\circ}$ NGP-cap and SGP-cap, 
respectively; while in
panels (c) and (d) we show the ACS in the $60^{\circ}$ NGP-cap and SGP-cap, 
respectively.
In Figure~\ref{figure4}, we also show four plots:
panels (a) and (b) reveal the ACS in the $90^{\circ}$-cap, outside the Kp2
mask, in the Northern and Southern Galactic hemispheres (NGH, SGH), 
respectively;
panels (c) and (d) show the ACS in the $90^{\circ}$-cap, outside the Kp2 
mask, in the NGH and SGH, respectively, but this time we removed the 
quadrupole component of the CMB maps.

Some interesting features in these plots are evident.
The first noticeable result is that, in each one of these figures the 
ACS corresponding to the three frequency maps Q, V, and W are practically 
indistinguishable.
It is well known (see Bennett et al.~\cite{WMAPb}) that foregrounds have a 
frequency-band dependence, thus when incorrectly subtracted it is expected 
to show up at a different level in the frequency maps Q, V, and W.
To investigate the robustness of our results with respect to a different 
sky cut, we computed the ACS for the NGH and SGH (not shown here) for data 
outside the more severe galactic cut given by the WMAP Kp0 mask (it cuts 
$\sim 25 \%$ of the total data, while the Kp2 mask cuts $\sim 15 \%$),
where we found similar results.
Thus, one concludes that the ACS found in these maps are remarkably stable
with respect to both frequency bands and sky cuts, and seem unlikely to be
compromised by residual foregrounds.

The second interesting fact is the net asymmetry between the 
$90^{\circ}$-caps in NGH and SGH: 
the Northern hemisphere is almost structureless 
(the variance's square-root 
for this ACS is $\langle \sigma_N^{\mbox{\small Q,V,W}} \rangle=0.0054$ 
while $\sigma^{\mbox{\rm\small Expected}}=$ $0.012 \pm 0.006$), 
instead the ACS in the Southern hemisphere are relatively intense 
($\langle \sigma_S^{\mbox{\small Q,V,W}} \rangle = 0.016$ 
while $\sigma^{\mbox{\rm\small Expected}}=$ $0.015 \pm 0.008$).
In this form, the South/North ratio 
$\sigma_{S/N} \equiv \sigma_{S} / \sigma_{N} = 2.97$
quantifies the evident SGH/NGH asymmetry observed in the plots of 
figure~\ref{figure4}. 
The statistical \,significance of this result are evaluated in the next 
section, by means of the Monte Carlo maps, in terms of the standard 
covariance matrix $\chi^2$ statistic. 

$\!$Finally, $\!$in order to check the possibility that the low value 
of the quadrupole moment could be \mbox{responsible} for this peculiar 
ACS behavior, we removed the quadrupole $\!$component, $\!$after applying 
the $\!$Kp2 mask, in all the six maps, and then compute again the 
MPASH-minus-EPASH for data in the $90^{\circ}$-caps in the NGH and SGH. 
\mbox{Our results can be seen in the Figures \ref{figure4}c and \ref{figure4}d, 
where} this time 
\mbox{$\langle \sigma_N^{\mbox{\small Q,V,W}} \rangle=0.0065$ in 
the NGH, and $\langle \sigma_S^{\mbox{\small Q,V,W}} \rangle$} $=0.019$ 
in the SGH. 
The large value now obtained for the ratio $\sigma_{S/N}\!=\!2.96$, 
confirms that the North/South asymmetry is not related to the 
low quadrupole value.

A natural question concerns the frequency that such a ratio is
$\sigma_{S/N} \simeq 3$ ($\sigma_{S/N} = 3.2$ if data from Coadded map 
is included in the average) is present in the ensemble of Monte Carlo 
maps.
Analysing the MPASH-minus-EPASH functions for similar NGH and SGH 
$90^{\circ}$-caps in 1\,000 Monte Carlo maps, we found that the 
corresponding values of $\,\sigma_{S/N} \in [0.2,5.4]$, 
with mean $\sigma_{S/N}^{\mbox{\rm\small mean}} = 1.1 \pm 0.7$. 
A close inspection of the values in such interval results in that 
less than 2\% of them satisfies $\sigma_{S/N} > 3$.

\section{Statistical analyses}

In this section we present the statistical analysis of the ACS obtained 
from WMAP data against analogous results derived from the 1\,000 Monte 
Carlo CMB maps. 
The degree of agreement between the simulations and the observations are 
quantified in terms of a standard covariance matrix $\chi^2$ statistic, 
including all bin-to-bin correlations, 
\vspace{-0.2cm}
\begin{eqnarray}
\chi^2 = \sum_{i,j=1}^{N_{\rm{bins}}} ( f_i - \langle f_i \rangle ) M^{-1}_{ij} 
                   ( f_j - \langle f_j \rangle ) \, , \nonumber
\end{eqnarray}

\vspace{-0.2cm}
\noindent
where $\langle f_i \rangle$ is the mean of the MPASH-minus-EPASH functions 
as determined by the Monte Carlo maps, and 
$M_{ij} \equiv \langle f_i f_j \rangle - \langle f_i \rangle \langle f_j \rangle$ 
is the covariance matrix.
The results from these computations, divided in three groups for the 
$45^{\circ}$-, $60^{\circ}$-, and $90^{\circ}$-caps, respectively, 
are reported in Table 1.
The first column indicates the CMB map investigated, the second and third 
ones indicate the frequencies of Monte Carlo simulations with a lower 
$\chi^2$ value than the WMAP data for the NGH and SGH, respectively.
The last column indicates the frequency of simulations with a smaller 
$\chi^2_{\mbox{\rm\small SGP}} / \chi^2_{\mbox{\rm\small NGP}}$.
These results determine a parameter, the ratio of the $\chi^2$ values for 
the NGH and SGH, that is used to quantify the degree of asymmetry 
between both hemispheres (Eriksen et al.~\cite{Eriksen04a},~\cite{Eriksen05a}).

The results reported in Table 1 are as follows. 
In all the cases, the NGH and SGH show, indepedently, acceptable values 
of probability (frequency of simulated maps,  given the model, where the 
MPASH-minus-EPASH function of such maps has a lower $\chi^2$ value than 
the WMAP data). 
However, data in the third column says that the antipodal $90^{\circ}$-caps 
are marginally consistent internally (when considered simultaneously one as 
opposite to the other): 
for the Q, V, W, and Coadded maps the mean of the $\chi^2_{S/N}$ values is 
$\ga 0.98$, or equivalently $\chi^2_{N/S} \la 0.02$, which means that 
$\la 2\%$ of the simulations have a smaller ratio of $\chi^2$.
It is worth to notice that these results appearing in the $90^{\circ}$-caps 
are shown to be robust with respect to frequency (the same result for Q, V, 
W, and Coadded CMB maps).
Therefore, we conclude that there is a significant discrepancy, at the 
$\ga 98\%$ CL, 
between WMAP data and the ACS expected in a statistically isotropic Universe.

\begin{table}
\caption[]{$\chi^2$-test for the ACS in different caps.
The first column indicates the CMB map investigated, the second and third 
ones indicate the frequency of Monte Carlo simulations with a lower 
$\chi^2$ value than the WMAP data.
The last column indicates the frequency of simulations with a smaller 
$\chi^2_{\mbox{\rm SGH}}/\chi^2_{\mbox{\rm NGH}}$.}
\vspace{-0.75cm}
$$ 
         \begin{array}{p{0.55\linewidth}lll}
         \hline \hline 
         \noalign{\smallskip}
CMB map & \hspace{-0.8cm} \chi^2\!\!-\!\mbox{\rm NGH} \, & \!\chi^2\!\!-\!\mbox{\rm SGH} 
& \chi^2_{S/N} \\
            \noalign{\smallskip}
            \hline
            \noalign{\smallskip}
\hspace{2.5cm} $45^{\circ}$  &      &      &     \\
             Q    & \hspace{-0.5cm} 0.914 &  0.891 &  0.499   \\
             V    & \hspace{-0.5cm} 0.854 &  0.902 &  0.624   \\
             W    & \hspace{-0.5cm} 0.654 &  0.782 &  0.616   \\
    {\sc Coadded} & \hspace{-0.5cm} 0.969 &  0.967 &  0.548   \\
             TOH  & \hspace{-0.5cm} 0.865 &  0.759 &  0.419   \\
             LILC & \hspace{-0.5cm} 0.817 &  0.882 &  0.626   \\
\hline 
\hspace{2.5cm} $60^{\circ}$   &     &      &   \\ 
             Q    & \hspace{-0.5cm} 0.715 &  0.971 &  0.949  \\
             V    & \hspace{-0.5cm} 0.637 &  0.982 &  0.956  \\
             W    & \hspace{-0.5cm} 0.361 &  0.959 &  0.930  \\
    {\sc Coadded} & \hspace{-0.5cm} 0.550 &  0.909 &  0.822  \\
             TOH  & \hspace{-0.5cm} 0.786 &  0.926 &  0.712  \\
             LILC & \hspace{-0.5cm} 0.154 &  0.884 &  0.891  \\
\hline 
\hspace{2.5cm} $90^{\circ}$   &     &      &   \\
             Q    & \hspace{-0.5cm} 0.276 &  0.959 &  0.985  \\
             V    & \hspace{-0.5cm} 0.243 &  0.810 &  0.965  \\
             W    & \hspace{-0.5cm} 0.278 &  0.968 &  0.976  \\
    {\sc Coadded} & \hspace{-0.5cm} 0.143 &  0.929 &  0.990  \\
             TOH  & \hspace{-0.5cm} 0.164 &  0.724 &  0.846  \\
             LILC & \hspace{-0.5cm} 0.176 &  0.770 &  0.853  \\
             \noalign{\smallskip}
             \hline\hline
             \end{array}
     $$ 
\vspace{-0.8cm}
\end{table}

\section{Conclusions} 

Here we have shown through the full-sky analyses of the ACS of the WMAP 
data (outside the Kp0 or Kp2 sky cuts) that the `anomalous' lack of 
power at large angular scales is mainly explained by the low 
quadrupole-moment value (see Efstathiou~\cite{GE}, Gazta\~naga et 
al.~\cite{Gaztanaga}, and Slosar et al.~\cite{Slosar} for more detailed 
discussions; see also Bielewicz et al.~\cite{Bielewicz04} for complete 
calculations, in particular in similar band maps as here). 
We confirmed this result by analysing also the WMAP data after removing 
the quadrupole component, obtaining a better (but not exactly) agreement 
with what is expected in a $\Lambda$CDM model with zero quadrupole.

We also performed partial-sky coverage analyses and showed that there is 
significative evidence, at the $98\%-99\%$ CL, of an anomalous 
distribution of the CMB power in opposite Galactic hemispheres.
In fact, in less than 2\% of the Monte Carlo CMB maps analyzed, data in 
opposite hemispheres present analogous bin-to-bin correlations as those 
shown by the $90^{\circ}$-caps with WMAP data. 
More interestingly, the excess of correlations in the comparison of 
opposite antipodal caps is manifestly observed in the South/North 
$90^{\circ}$-caps, but it is not seen neither in the examination of 
the antipodal $45^{\circ}$- or $60^{\circ}$-caps.

The fact that the South/North asymmetry persists at a significant level 
in several frequency bands (the Q, V, and W CMB maps) and with different 
sky cuts (the Kp0 and Kp2 masks) excludes a simple explanation based on 
systematic effects and foregrounds (see Hinshaw et al. \cite{Hinshaw03a}; 
Eriksen et al. \cite{Eriksen04a},~\cite{Eriksen04b}; Hansen et 
al.~\cite{Hansen04b}). 

\begin{acknowledgements}
We acknowledge use of the Legacy Archive for Microwave Background 
Data Analysis (LAMBDA), and use of the TOH map (Tegmark et 
al.~\cite{TOH}). 
Thanks to M.J. Rebou\c{c}as, K. Land, and P. Bielewicz for useful 
comments. 
We are grateful to H.K. Eriksen for valuable suggestions.
T.V. and C.A.W. acknowledge CNPq grants 305219/2004-9 and 307433/2004-8, 
respectively, and a FAPESP grant \mbox{00/06770-2}. 
R.L. and I.F. thank CAPES fellowships, and A.B. thanks a PCI/7B-CNPq 
fellowship. 
Some of the results in this paper were derived using the 
HEALPix package (G\'orski et al.~\cite{Gorski}).
\end{acknowledgements}


\end{document}